\documentclass[a4paper, 12pt]{article}
\usepackage{amsmath,amsfonts,amssymb,epsfig,psfrag,subfigure}
\usepackage{amsmath,amsfonts,amssymb}
 
\renewcommand{\le}{\leqslant}
\renewcommand{\ge}{\geqslant}

\newcommand{\ii}{\textrm{i}}
\newcommand{\ee}{\textrm{e}}
\renewcommand{\vec}[1]{\boldsymbol{#1}}
\newcommand{\be}{\begin{equation}}
\newcommand{\en}{\end{equation}}

\begin{document}

\title{Seismic Rayleigh waves on an \\exponentially graded, orthotropic
half-space}
\author{Michel Destrade}
\date{2006}
\maketitle

\begin{abstract}

Efforts at modelling the propagation of seismic waves in half-spaces 
with continuously varying properties have been mostly focused 
on shear-horizontal waves.
Here a sagittaly polarized (Rayleigh type) wave travels along a 
symmetry axis (and is attenuated along another) of an orthotropic 
material with stiffnesses and mass density varying in the same 
exponential manner with depth. 
Contrary to what could be expected at {f}irst sight, the analysis is 
very similar to that of the homogeneous half-space, with the main and 
capital difference that the Rayleigh wave is now dispersive. 
The results are illustrated numerically for 
(i) an orthotropic half-space typical of horizontally layered and 
vertically fractured shales and (ii) for an isotropic half-space 
made of silica.
In both examples, the wave travels at a slower 
speed and penetrates deeper than in the homogeneous case; 
in the second example, the inhomogeneity can force the wave amplitude  
to oscillate as well as decay with depth, in marked contrast with the
homogeneous isotropic general case. 

\end{abstract}

\section{Introduction}

Love (1911) showed that a \textit{inhomogeneous} half-space, 
consisting of an elastic layer covering a semi-in{f}inite body made 
of a different elastic material, can sustain the propagation of a 
linearly polarized (shear horizontal) surface wave.
The Love wave is faster than the elliptically polarized (vertical) 
Rayleigh (1885) wave and it has been observed countless times during 
earthquakes or underground explosions. 
Another recorded phenomenon is that Rayleigh waves 
are \textit{dispersive}, a characteristic which is incompatible with 
the context of a homogeneous half-space given by Rayleigh (1885):
Love showed that his layer/substrate con{f}iguration could 
also support a two-partial, vertically polarized, surface wave. 
Because this con{f}iguration introduces a new characteristic 
length, the layer thickness $h$ (say), a dispersion 
parameter is now $k h$ where $k$ is the wave number, and that 
surface wave is dispersive.

Subsequent analyses introduced more and more layers to refine the 
model, until is was considered practical to view the inhomogeneity of 
the half-space as a \textit{continuous} variation of the material 
properties (Ewing \emph{et al.} 1957). 
Chief among these continuous variations is the one for which the 
elastic stiffnesses and the mass density vary exponentially with 
depth, all in the same manner, proportional to a common factor 
$\exp (-2 \alpha x_2)$ say, where $\alpha$ is the inverse of a 
inhomogeneity characteristic length, and $x_2$ is the coordinate along 
the normal to the free surface, so that here a dispersion 
parameter is now $\alpha/k$ for instance.
Hence Wilson (1942), Deresiewicz (1962), Dutta (1963), 
Bhattacharya (1970), and many others studied the 
propagation of surface waves in such inhomogeneous media;
they were however interested in shear-horizontal waves (Love-type).
The literature on Rayleigh-type surface waves in that type of media 
is quite scarce, probably because the dif{f}iculty exposed below is 
encountered quite early in the analysis.

In an anisotropic elastic body with continuously variable 
properties, the general equations of motion read
 \begin{equation} \label{motion1}
 C_{ijkl} u_{l,kl} + C_{ijkl,j} u_{l,k} 
   = \rho u_{i,tt},
 \end{equation}
where $\vec{u}$ is the mechanical displacement, 
and $C_{ijkl}$ and $\rho$ are the elastic stiffnesses and 
the mass density, respectively. 
Now consider the propagation of an inhomogeneous plane 
wave with speed $v$ and wave number $k$ in the $x_1$-direction,
and with attenuation in the $x_2$-direction,
 \begin{equation} \label{solnForm}
 \vec{u} = \vec{U^\circ} \ee^{\ii k(x_1 + q x_2 -v t)},
 \end{equation}
in a half-space $x_2 \ge 0$ made of an 
orthotropic\footnote[1]{An anisotropic material belongs to the
orthotropic symmetry class when it possesses three mutually orthogonal 
planes of mirror symmetry.}
material 
with an exponential depth pro{f}ile,
\begin{equation} \label{inhomogeneityRhombic}
 \{ c_{11}(x_2), c_{22}(x_2), c_{12}(x_2), 
     c_{66}(x_2), \rho(x_2) \}  
  =  \ee^{- 2\alpha x_2}
       \{ c^\circ_{11}, c^\circ_{22}, c^\circ_{12}, 
        c^\circ_{66}, \rho^\circ \}.
\end{equation}
Here the $x_1$, $x_2$, $x_3$ directions are aligned 
with the axes of symmetry, $\alpha$ is a real number, 
and the $c^\circ_{ij}$ and $\rho^\circ$ are constants;
also, $\vec{U^\circ}$ is a constant vector and $q$ a 
complex number so that the attenuation factor is $k \Im(q)$.
Then the equations of motion  \eqref{motion1} yield 
\begin{equation} \label{quartic}
 \begin{bmatrix}
  c_{66}^\circ q^2 + c_{11}^\circ - \rho v^2 
   + 2\ii (\alpha/k) q c_{66}^\circ 
   & q(c_{12}^\circ + c_{66}^\circ) 
    + 2\ii (\alpha/k) c_{66}^\circ 
\\
   q(c_{12}^\circ + c_{66}^\circ) 
    + 2\ii (\alpha/k) c_{12}^\circ  
   &   c_{22}^\circ q^2 + c_{66}^\circ - \rho v^2 
   + 2\ii (\alpha/k) q c_{22}^\circ
 \end{bmatrix}
 \vec{U^\circ}
  = \vec{0}.
\end{equation} 
At $\alpha = 0$, the material is \emph{homogeneous}, 
and the associated determinantal equation 
-- the propagation condition -- is a 
real \emph{quadratic} in $q^2$ which can be solved exactly
(Sveklo, 1948). 

At $\alpha \ne 0$, the propagation condition is 
seemingly a \emph{quartic} in $q$ with complex coef{f}icients, 
whose analytical resolution might appear to be a daunting 
task and to preclude further progress toward the 
completion of a boundary value problem
(note that it remains a quartic even when the material is isotropic.)
Hence, Das \textit{et al.} (1992) and Pal \& Acharya (1998) 
stopped their analytical study of that problem at that very point.
In fact the transformation of the quartic to its 
canonical form reveals that it is a \emph{quadratic}
in $q + \ii (\alpha / k)$, with \emph{real} coef{f}icients. 
That this is so has rarely been identi{f}ied: 
Biot (1965), in the context of incremental static deformations,
seems to be the only one who has recognized this simpli{f}ication. 
The present paper shows that the Stroh (1962) 
formulation of this problem, combined with a change of 
unknown functions,  leads naturally to the biquadratic in question.
Then the propagation condition can be solved exactly, and the general 
solution of form \eqref{solnForm} to the equations of motion follows. 
In particular, the resolution of the dispersive Rayleigh wave boundary 
value problem poses no particular dif{f}iculty after all. 
Section 2 exposes this analysis, and Section 3 applies it to two 
types of exponentially graded half-spaces:
one which would be made of orthotropic shales if 
$\alpha \rightarrow 0$ and another which would be made of silica 
(isotropic).
There, it is seen for both examples that the influence of the 
inhomogeneity is more marked upon the wave speed (rapidly decreasing 
with $\alpha/k$) than upon the attenuation factors (slowing increasing 
with $\alpha/k$).
It is also found that the attenuation factors for the displacement 
amplitudes are \emph{distinct} from those for the traction
amplitudes, and that the amplitudes can decay in an \emph{oscillating} 
manner for the isotropic silica. 
These two features are unusual and are clearly due to the 
inhomogeneity.

The overall aim of the paper is to show that simple, analytical, 
exact results can be obtained for seismic Rayleigh wave propagation in 
an anisotropic, inhomogeneous Earth. 
Of course it is unlikely any ``real'' inhomogeneity can be such that 
the stiffnesses and the mass density all vary in the same manner as in 
\eqref{inhomogeneityRhombic}, because it then leads to bulk wave 
speeds (proportional to the square root of stiffnesses divided by the 
density) which are constant with depth. 
The analysis of more realistic models must turn to numerical 
simulations such as those based on the {f}inite difference technique 
or on the pseudospectral technique or on techniques with Fourier or 
other function expansions (e.g. Tessmer 1995).
These methods however encounter dif{f}iculties for the implementation 
of accurate boundary conditions and of strong heterogeneity. 
The spectral element method seem to alleviate those dif{f}iculties 
but, as stressed by Komatitsch \& al. (2000), it must be validated 
against analytical solutions. 
Such a solution validation procedure is indeed a crucial necessity of 
numerical simulations in geophysics, where different software 
packages can give widely different predictions (Hatton 1997).

\section{The dispersion equation}

Consider the propagation of a Rayleigh wave,
traveling with speed $v$ and wave number $k$ in the $x_1$-direction,
in an inhomogeneous half-space $x_2 \ge 0$ made of the orthotropic 
material presented in the Introduction. 
The associated mechanical quantities are the displacement components 
$u_j$ and the traction components $\sigma_{j2}$ ($j = 1,2$). 
They are now taken in the form
\begin{equation} \label{wave}
\{ u_j, \sigma_{j2} \}(x_1, x_2, t)
  = \{ U_j(x_2),  \ii t_{j2}(x_2)\}
                \ee^{\ii k(x_1 - vt)},
\end{equation}
where the $U_j$, $t_{j2}$ ($j = 1,2$) are yet unknown 
functions of $x_2$ alone, to be determined from the 
equations of motion and from the boundary conditions.

The equations of motion:
$\sigma_{ij,j} = \rho u_{i,tt}$, can be written as 
the second-order differential system \eqref{motion1},
or as the following {f}irst-order differential system,
\begin{equation} \label{motion2}
   \begin{bmatrix} \vec{U}' \\ \vec{t}' \end{bmatrix}
     = \ii \begin{bmatrix}
                    k N_1 & \ee^{2 \alpha x_2} N_2 \\
                     k^2 \ee^{- 2\alpha x_2} K  & k N_1^t
                             \end{bmatrix}
   \begin{bmatrix} \vec{U} \\ \vec{t} \end{bmatrix}.
\end{equation}
Here $N_1$, $N_2$, $K$ are the usual constant matrices
of Stroh (1962), given by 
\be
 N_1 = 
\begin{bmatrix}
    0           & -1 \\
    -\dfrac{c^\circ_{12}}{c^\circ_{22}}&  0
  \end{bmatrix},
\quad
 N_2 = 
\begin{bmatrix}
  \dfrac{1}{c^\circ_{66}} &  0  \\
      0   & \dfrac{1}{c^\circ_{22}}
\end{bmatrix},
\quad
 K = 
\begin{bmatrix}
  X -  c^\circ &  0  \\
      0        & X
\end{bmatrix}.
\en
where 
$c^\circ := c^\circ_{11} - \dfrac{c^{\circ 2}_{12}}{c^\circ_{22}}$ 
and $X := \rho^\circ v^2$.
With the new vector function $\vec{\xi}$, 
de{f}ined as
\begin{equation} \label{scaling}
\vec{\xi} (x_2) := 
 [\ee^{- \alpha x_2}\vec{U}(x_2),\ee^{ \alpha x_2}\vec{t}(x_2)]^t,
\end{equation}
the system  \eqref{motion2} becomes
\begin{equation} \label{scaledSystem}
   \vec{\xi}' 
     = \ii k N  \vec{\xi}
\quad\text{ where } 
\quad
  N := \begin{bmatrix}
       N_1 + \ii (\alpha/k) I & (1/k) N_2 \\
       k K  & N_1^t - \ii (\alpha/k) I
       \end{bmatrix}.
 \end{equation}
Hence the apparently anodyne change of unknown functions 
\eqref{scaling} transforms the differential system with 
variable coef{f}icients \eqref{motion2} into one with 
constant coef{f}icients. 

Now solve the differential system 
 \eqref{scaledSystem}
with a solution in exponential evanescent form,
\begin{equation} \label{inequality}
 \vec{\xi}(x_2) = \text{e}^{i k p x_2} \vec{\zeta}, 
 \quad 
 \Im(p) > |\alpha|/k,
\end{equation}
where $\vec{\zeta}$ is a constant vector, $p$ is a scalar,
and the inequality ensures that 
\be
\vec{u}(\infty) = \vec{0}, \quad 
\vec{t}(\infty) = \vec{0}, \quad 
\vec{\xi}(\infty) = \vec{0},
\en
because by  \eqref{scaling} and  \eqref{inequality}$_1$, 
$\vec{u}(x_2)$ behaves as: $\exp k(\ii p + \alpha/k)x_2$ and 
$\vec{t}(x_2)$ behaves as: $\exp k(\ii p - \alpha/k)x_2$.
Note in passing that, in sharp contrast to the homogeneous case, 
the displacement {f}ield and the traction {f}ield have 
\emph{different} attenuation factors: 
for $\vec{u}$ it is $k[\Im(p) - \alpha/k]$;  
for $\vec{t}$ it is $k[\Im(p) + \alpha/k]$.

Then $\vec{\zeta}$  and $p$ are solutions to the eigenvalue 
problem: $N\vec{\zeta} = p \vec{\zeta}$.
The associated determinantal equation is the 
\textit{propagation condition}, here a \textit{biquadratic} 
(and not a quartic as Eq.\eqref{quartic} suggested),
\begin{equation} \label{propCond}
 p^4 - S p^2 + P =0,
\end{equation}
where
\begin{align} \label{P&S}
 & S = [c^{\circ 2}_{12} + 2c^\circ_{12}c^\circ_{66} 
        - c^\circ_{11}c^\circ_{22}
         + (c^\circ_{22} + c^\circ_{66})X]/(c^\circ_{22}c^\circ_{66})
          - 2(\alpha/k)^2,
\notag \\
 & P =  (c^\circ_{11}-X)(c^\circ_{66} - X)/(c^\circ_{22}c^\circ_{66})
  \notag \\
 & \phantom{123456}
    - (\alpha/k)^2
     [c^{\circ 2}_{12} - 2c^\circ_{12}c^\circ_{66} 
        - c^\circ_{11}c^\circ_{22}
         + (c^\circ_{22} + c^\circ_{66})X]/(c^\circ_{22}c^\circ_{66})
 \notag \\
 & \phantom{12345678910}
    + (\alpha/k)^4.
\end{align} 
Let $p_1$ and $p_2$ be the two roots of  \eqref{propCond} satisfying 
inequality \eqref{inequality}. 
That pair may be in one of the two forms: 
$p_1 = ib_1$, $p_2 = ib_2$, or 
$p_1 = -a + ib$, $p_2 = a + ib$, where $b$, $b_1$, $b_2$ are 
positive.
In both cases, $p_1 p_2$ is a real negative number and $p_1 + p_2$ is 
a purely imaginary number with positive imaginary part.
It follows in turn that 
\be \label{p&s}
 p_1 p_2 = -\sqrt{p_1^2 p_2^2} = -\sqrt{P}, \quad 
 p_1 + p_2 = \ii \sqrt{-(p_1 + p_2)^2}
           = \ii\sqrt{ 2\sqrt{P} - S}.
\en
The associated eigenvectors $\vec{\zeta^1}$, $\vec{\zeta^2}$ are 
determined from: $N\vec{\zeta^j} = p_j \vec{\zeta^j}$,  as
\be 
 \vec{\zeta^j} = 
 \begin{bmatrix} 
  p_j^2 + 2 \ii (\alpha/k)p_j - e_0
\vspace{4pt}\\
   -[p_j^3 + \ii (\alpha/k)p_j + f_1 p_j + \ii (\alpha/k)f_0]
\vspace{4pt}\\
-k[g_1 p_j + \ii (\alpha/k)g_0]
\vspace{4pt}\\
-k[X p_j^2 + h_0]
\end{bmatrix},
\en
where the non-dimensional quantities $e_0$, $f_1$, $f_0$ appearing 
in the displacement components are given by
\begin{align}
& e_0 = (\alpha/k)^2 
     + c^\circ_{12}(c^\circ_{66} - X)/(c^\circ_{22}c^\circ_{66}),
\notag \\
& f_1 =  (\alpha/k)^2 
          + (c^\circ - X)/c^\circ_{66} - c^\circ_{12}/c^\circ_{22},
\notag \\
& f_0 =  (\alpha/k)^2 
          + (c^\circ - X)/c^\circ_{66} + c^\circ_{12}/c^\circ_{22},
\end{align}
and the quantities $g_1$, $g_0$, $h_0$ (dimensions of a stiffness) 
appearing in the traction components are given by
\begin{align} \label{g1g0h0}
& g_1 = c^\circ - (1 + c^\circ_{12}/c^\circ_{22})X,
\notag \\
& g_0 = c^\circ - (1 - c^\circ_{12}/c^\circ_{22})X,
\notag \\
& h_0 =  (\alpha/k)^2 X
          - (c^\circ - X)(c^\circ_{66} - X)/c^\circ_{66}.
\end{align}
Now construct the general solution to the equations of motion 
\eqref{scaledSystem} as
\be \label{gnlSln}
 \vec{\xi}(x_2) = 
  \gamma_1 \text{e}^{i k p_1 x_2} \vec{\zeta^1} 
  +  \gamma_2 \text{e}^{i k p_2 x_2} \vec{\zeta^2},
\en
where the constants $\gamma_1$, $\gamma_2$ are such that the surface 
$x_2=0$ is free of tractions: $\vec{t}(0) = \vec{0}$ or equivalently: 
$\vec{\xi}(0) = [\vec{U}(0), \vec{0}]^t$.
This condition leads to a homogeneous linear system of two equations 
for the two constants, whose determinant must be zero. 
After factorization and use of \eqref{p&s}, the dispersion equation
follows as
\be \label{dispersion}
g_1 (X \sqrt{P} + h_0) + (\alpha/k) g_0 X \sqrt{ 2\sqrt{P} - S} = 0.
\en

This equation is fully explicit ($X$ is the sole unknown) because 
$P$ and $S$ are given in \eqref{P&S} and $g_1$, $g_0$, $h_0$ 
are given in \eqref{g1g0h0}, and it is clearly \textit{dispersive} 
due to the multiple appearance of the dispersion parameter 
$\alpha/k$.
At $\alpha = 0$ (homogeneous substrate), it simpli{f}ies
to 
\be \label{secular}
X\sqrt{\dfrac{(c^\circ_{11} - X)(c^\circ_{66} - X)}
                 {c^\circ_{22} c^\circ_{66}}}
 - \dfrac{(c^\circ - X)(c^\circ_{66} - X)}
                 {c^\circ_{66}} = 0,
\en
the classic (non-dispersive) secular equation for Rayleigh waves in 
orthotropic solids.

\section{Examples: exponentially graded shales and silica}

As two examples of application, consider in turn that the half-space 
is made of a material with exponentially variable properties 
which is \textit{(i)} with orthotropic symmetry and \textit{(ii)}
isotropic.

In Example \textit{(i)} the starting point is a model proposed by 
Schoenberg and Helbig (1997), accounting for the vertical {f}ine 
strati{f}ication and the vertical fractures found in many shales. 
In their numerical simulations, they used the following 
orthotropic  elastic stiffness matrix,
\be \label{shales}
[c^\circ_{ij}] = \rho^\circ
   \begin{bmatrix}
  9    & 3.6  & 2.25   & 0 & 0   & 0     \\
  3.6  & 9.89 & 2.4    & 0 & 0   & 0     \\
  2.25 & 2.4  & 5.9375 & 0 & 0   & 0     \\
  0    & 0    & 0      & 2 & 0   & 0     \\
  0    & 0    & 0      & 0 & 1.6 & 0     \\
  0    & 0    & 0      & 0 & 0   & 2.182 
    \end{bmatrix}.
\en
Note that here the matrix is density-normalized so that its components 
have the dimensions of squared speeds, expressed in (km/s)$^2$ (Schoenberg 1994). 
Schoenberg and Helbig remark that ``the rock mass behaves as if it 
contains systems of parallel fractures increasing the compliance in
some directions''; 
integrating this information, $\alpha$ is assumed positive here. 
Also, \eqref{shales} is assumed to be the elastic stiffness matrix 
on the free surface $x_2 = 0$. 

In Example \textit{(ii)}, the half-space is assumed  to be made of an 
exponentially graded material such that at the boundary, 
$c^\circ_{11} = 7.85$, $c^\circ_{12} = 1.61$ ($10^{10}$ N/m$^2$) 
and $\rho^\circ = 2203$ kg/m$^3$ as in silica 
(Royer \& Dieulesaint 2000).
Here too, $\alpha$ is taken positive.

If the half-spaces were homogeneous, then the Rayleigh wave would 
travel with speed $v^\circ = \sqrt{X/\rho^\circ}$ 
where $X$ is given by \eqref{secular}, that is 
$v^\circ = 1.412$ km/s for shales and $v^\circ = 3409$ m/s for silica.
For any given dispersion parameter $\alpha / k$, the dispersion 
equation \eqref{dispersion} in the inhomogeneous half-spaces 
gives a unique root $X$. 
In both examples, it has then been checked that for that $X$, 
the propagation condition \eqref{propCond}  gives two roots such that 
the inequality \eqref{inequality}$_2$ is always satis{f}ied. 
Thus the surface wave exists for arbitrary value of $\alpha / k$, 
and it travels with speed $v = \sqrt{X/\rho^\circ}$.
Although this state of affair is acceptable mathematically, it seems 
reasonable to limit the range of $\alpha / k$ to values where the 
wave amplitude decays faster than the inhomogeneity. 
Because the amplitudes of the tractions  
$\vec{t}$ decay as $\exp -k[\Im(p) + \alpha/k]$, 
they always decrease faster than $\exp -2\alpha x_2$ by 
\eqref{inequality}$_2$; 
on the other hand, the amplitudes of the displacements 
$\vec{u}$ decay as $\exp -k[\Im(p) - \alpha/k]$: 
thus they decrease faster than the 
inhomogeneity as long as $\Im(p) > 3\alpha/k$. 
In Example \textit{(i)}, it turns out that this latter inequality is 
veri{f}ied for $\alpha / k < 0.107$, and in Example \textit{(ii)}, 
for $\alpha / k < 0.274$.

Fig. 1 shows the variation of the wave speed (decreasing) and of 
$\Im(p_1)$, $\Im(p_2)$ (increasing) in Example\textit{(i)} over the 
range $0 \le \alpha/k \le 0.1$. 
It has also been checked there that the attenuation factors for both 
the displacements amplitudes ($k[\Im(p) - \alpha/k]$) and the tractions
amplitudes ($k[\Im(p) - \alpha/k]$) increase also. 
In conclusion, the surface wave travels at a slower speed in the 
inhomogeneous shales than in the homogeneous shales, and it is less 
localized.

Fig. 2 shows the variation of the wave speed (decreasing) and of 
$\Im(p_1)$, $\Im(p_2)$ (increasing) in Example\textit{(ii)} over the 
range $0 \le \alpha/k \le 0.25$. 
It has been checked again that the attenuation factors for both the 
displacements amplitudes ($k[\Im(p) - \alpha/k]$) and the tractions
amplitudes ($k[\Im(p) - \alpha/k]$) increase also. 
Here again, the surface wave travels at a noticeably slower speed in 
the inhomogeneous case than in the homogeneous case, and it is 
slightly less localized.
A most interesting phenomenon occurs at $\alpha / k \approx 0.211$ 
where the nature of the roots changes 
from the form: $p_1 = ib_1$, $p_2 = ib_2$, 
to the form: $p_1 = -a + ib$, $p_2 = a + ib$, so that the amplitudes 
switch from decaying in a real exponential manner to decaying in an 
exponential oscillating manner. 
This latter situation \textit{never} arises 
in a \textit{homogeneous} isotropic half-space.


\newpage 

\begin{figure}
 \centering 
  \mbox{\subfigure{\epsfig{figure=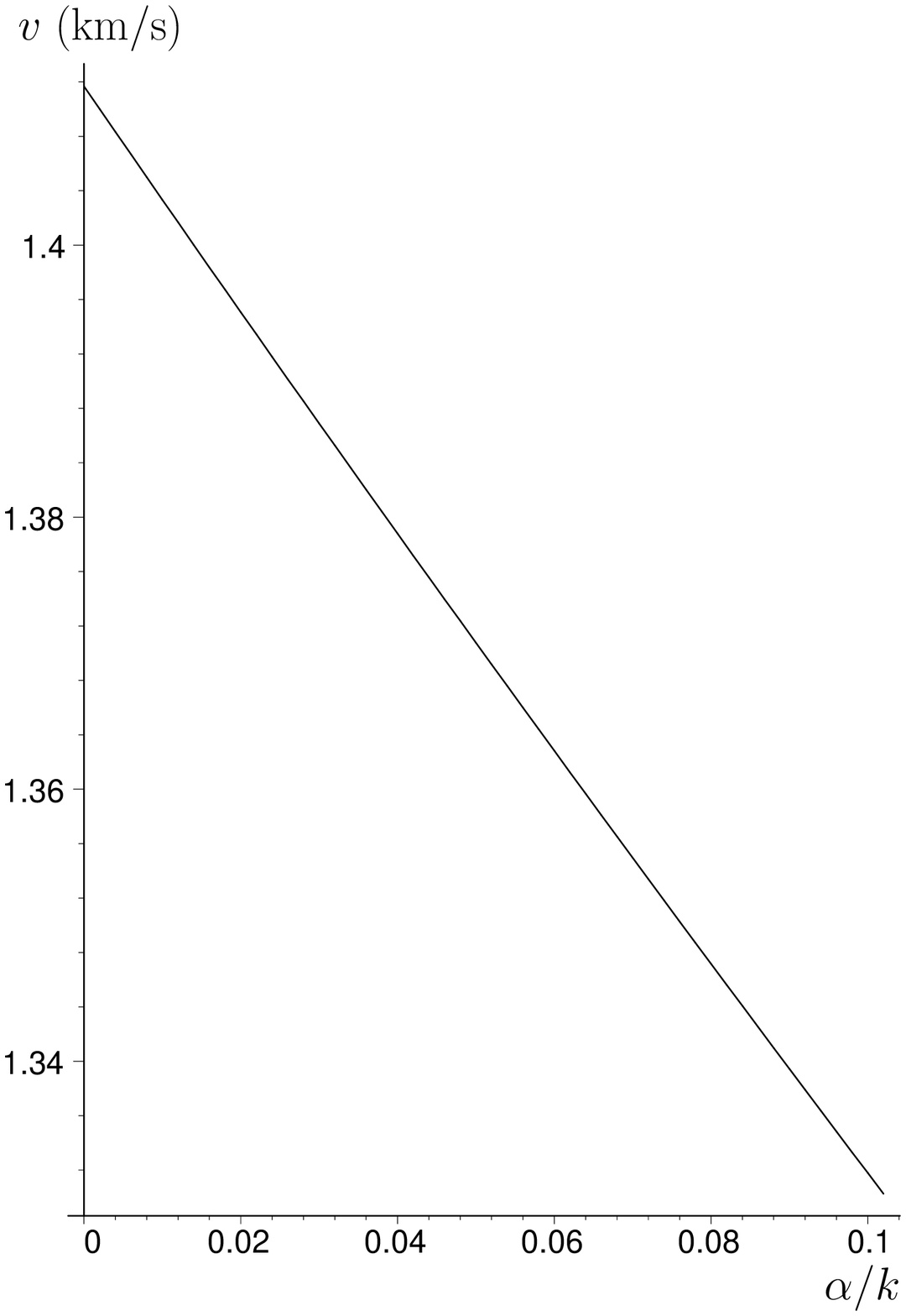, height=.45\textwidth,
width=.45\textwidth}}
  \quad \quad
     \subfigure{\epsfig{figure=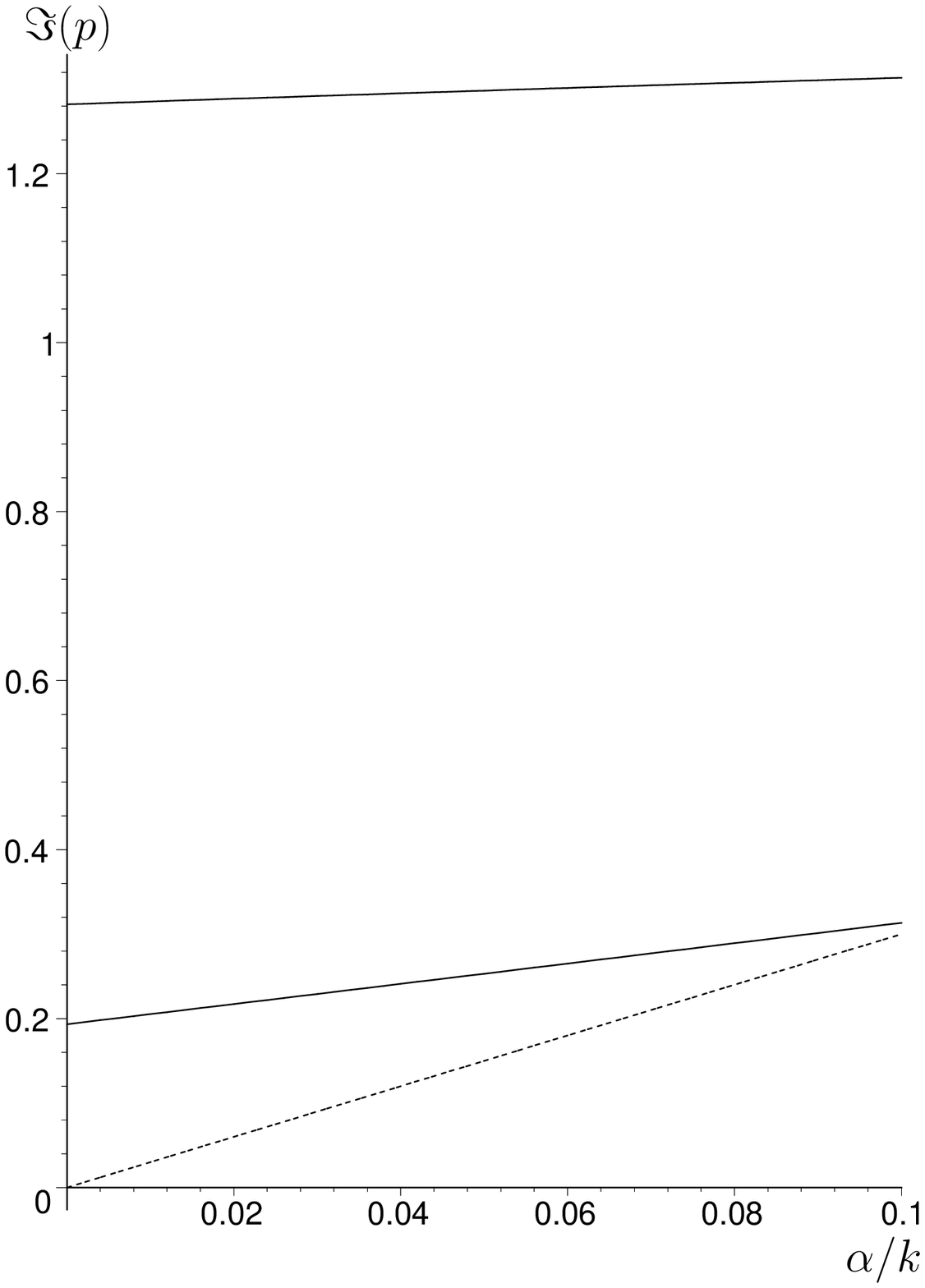, height=.45\textwidth,
width=.45\textwidth}}}
\caption{Exponentially graded orthotropic shales: 
variations with the dispersion parameter $\alpha/k$ of 
(a) the surface wave speed and (b) the imaginary parts of the 
quantities $p_1$ and $p_2$ appearing in 
\eqref{gnlSln} (the dashed line is the plot of $3\alpha/k$,
above which $\Im(p_1)$, $\Im(p_2)$ must be for the wave to decrease 
faster than the inhomogeneity).}
\end{figure}

\begin{figure}
 \centering 
  \mbox{\subfigure{\epsfig{figure=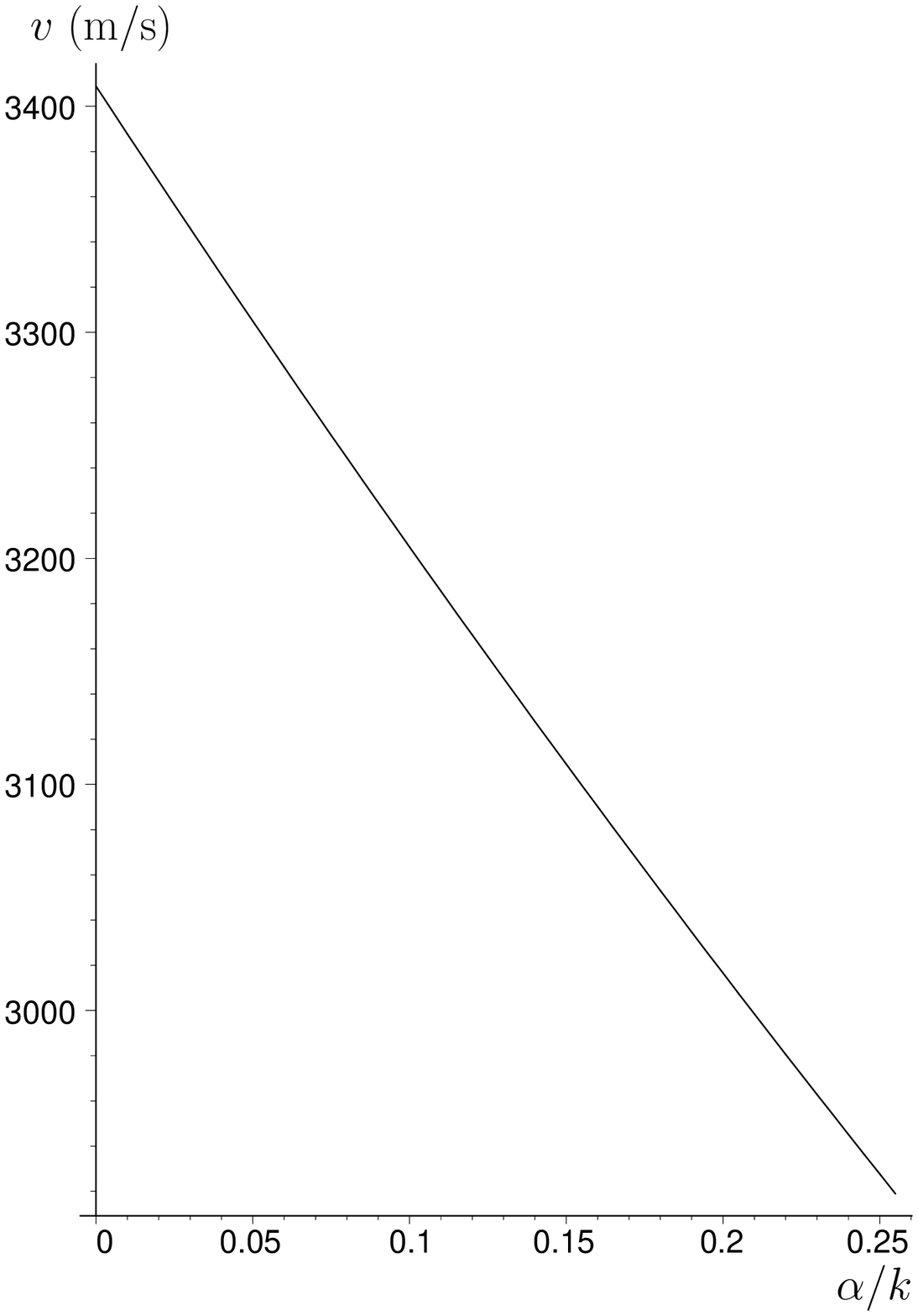, height=.45\textwidth,
width=.45\textwidth}}
  \quad \quad
     \subfigure{\epsfig{figure=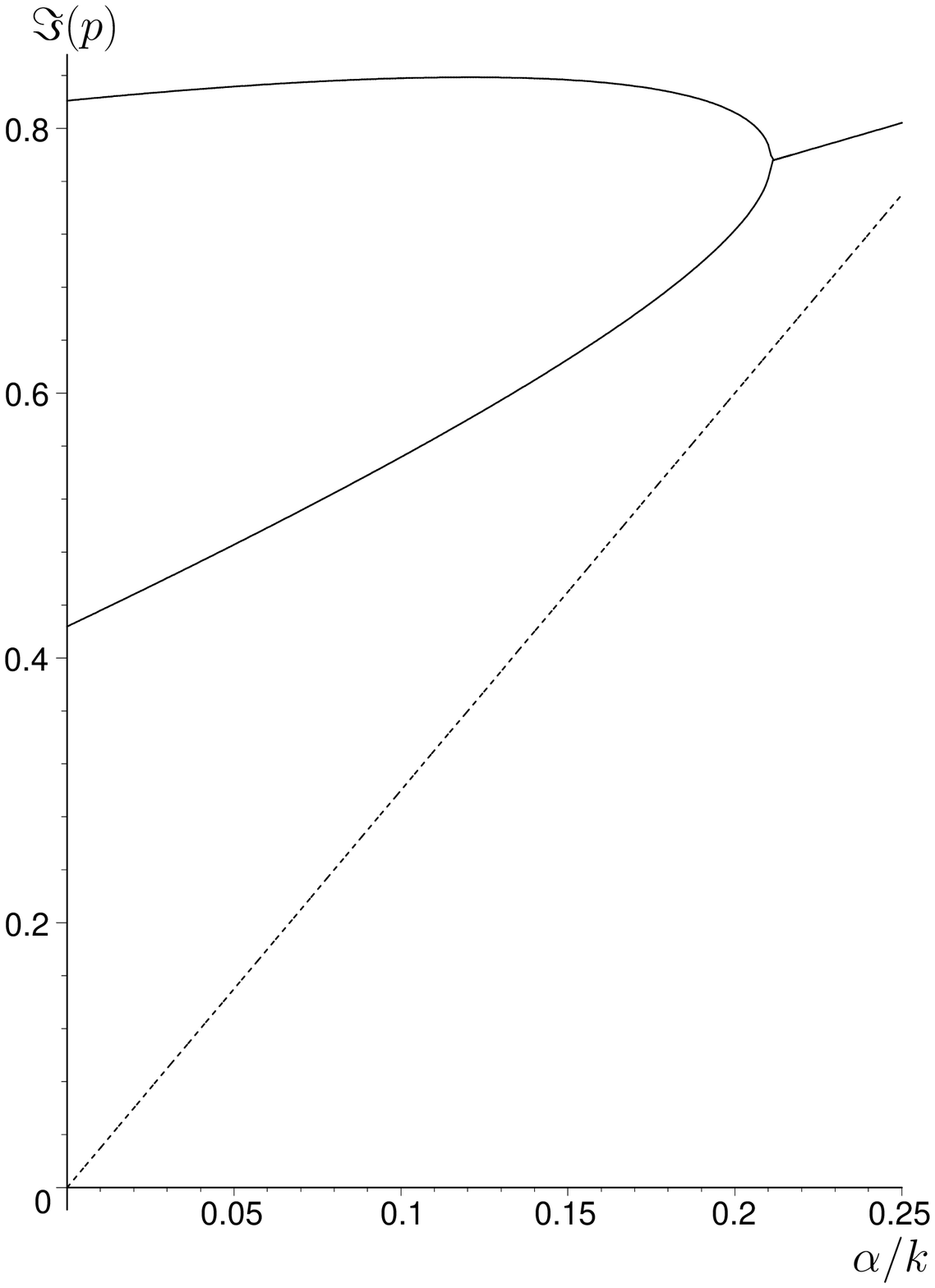, height=.45\textwidth,
width=.45\textwidth}}}
\caption{Exponentially graded silica: variations with the dispersion 
parameter $\alpha/k$ of (a) the surface wave speed and (b) the  
imaginary parts of the quantities $p_1$ and $p_2$ appearing in 
\eqref{gnlSln} (the dashed line is the plot of $3\alpha/k$).}
\end{figure}


\end{document}